\documentclass{aa501}
\usepackage{graphicx}

\begin{document}

\title{M2000 : an astrometric catalog in the Bordeaux Carte du Ciel zone $+11\degr \leq \delta \leq
+18\degr$\thanks{The catalog is distributed on CD-ROM and through the CDS 
(http://cdsweb.u-strasbg.fr/Cats.html).
Further information is given on the website http://www.observ.u-bordeaux.fr/$\sim$soubiran/m2000.htm}
}

\author{Rapaport M., Le Campion J.-F., Soubiran C., Daigne G., P\'eri\'e J.-P., Bosq F., Colin J.,  
Desbats J.-M., Ducourant C., Mazurier J.-M., Montignac G., Ralite N., R\'equi\`eme Y., Viateau B.}

\offprints{Michel Rapaport, rapaport@observ.u-bordeaux.fr}

\institute{Observatoire de Bordeaux, CNRS UMR 5804, BP 89, F-33270 Floirac, France}

\date{ }
\abstract{
During four years, systematic observations have been conducted in drift scan mode with the 
Bordeaux automated meridian circle in the declination band $+11\degr \leq \delta \leq +18\degr$. 
The resulting astrometric catalog
includes about $2.3\, 10^6$ stars down to the magnitude limit $V_M=16.3$. Nearly all stars (96\%) have been
observed at least 6 times, the catalog being complete down to $V_M=15.4$. The median internal standard errors 
in position is
$\sim$ 35 mas in the magnitude range $11 < V_M <
15$, which degrades to $\sim$ 50 mas when the faintest stars are considered. M2000 provides also one band 
photometry
with a median internal standard error of  $\sim$ 0.04 mag. Comparisons with the Hipparcos 
and bright part of Tycho-2 catalogs have enabled to estimate external errors in position to be lower 
than 40 mas. In this zone and at epoch 1998, the faint part of Tycho-2 is found to have an accuracy 
of 116 mas in 
$\alpha$ instead of 82 mas deduced from the model-based standard errors given in the catalog.
}
\authorrunning{Rapaport et al.}
\titlerunning{M2000}
\maketitle

\section{Introduction}
Accurate proper motions for large samples of stars are of a considerable interest for both the
maintenance and extension of optical reference frames and for the understanding of the structure 
and evolution of the Galaxy. At present 
the largest and most precise all-sky catalogs of proper motions are Hipparcos (ESA \cite{esa97}), very precise but
concerning only 120000 bright stars, and the recent Tycho-2
(H{\o}g et al. \cite{hog00}) which is complete down to V=11 with $2.5\, 10^6$ stars. In 2003, the USNO CCD 
Astrograph catalog (UCAC) will be available with an all-sky coverage down to V=16 and a precision similar
to Tycho-2 in positions and proper motions (Zacharias et al. \cite{zac00}). This will provide a valuable 
database
for galactic studies but until then there is a lack of large and complete samples of stars with 
accurate proper motions at intermediate magnitude 11-16.

The accuracy of proper motions depends on the accuracy of individual positions at different epochs 
and the time
baseline which separate them. An invaluable source of old epoch positions is the Carte du Ciel,
an international program conducted at the beginning of the century. It was the first 
photographic survey of the entire sky. In this
ambitious program, fields of $2.5\degr \times 2.5\degr$ were observed twice, to limiting blue magnitudes of 
$\sim$ 15 and $\sim$ 12.5 respectively. The bright part, the Astrographic Catalog (AC),
was entirely measured and produced a positional catalog of $4.6\, 10^6$ stars, the AC 2000 (Urban et al. 
\cite{urb98a}), which was combined to the recent Tycho catalog into the ACT catalog (Urban et
al. \cite{urb98b}) and which was also used to produce the Tycho-2 catalog.  While the AC positions
have been used to a large extend, it is not the case for the faint survey of the Carte du Ciel which
has only been used in punctual studies. 

As one of the participant of the Carte du Ciel, 
in charge of the declination band $+11\degr \leq \delta \leq +18\degr$, the Bordeaux Observatory owns 
511 Carte du Ciel plates covering this band. It was decided to digitize them in order to salvage 
this astronomical treasure before its deterioration and
to conduct at the same time a scientific program. The Bordeaux Observatory is also one of the last
urban observatories where astrometric observations are still possible thanks to its meridian circle 
which has been equiped with a CCD camera. Fully automatic and of a great stability, 
the Bordeaux meridian
circle has been proved to be very efficient with an internal precision better than 40 mas in 
both
coordinates, in the magnitude range $9 < V < 14 $ (Viateau et al. \cite{via99}). The M2000 program aims at
producing a 
catalog of positions and proper motions from the cross-identification between today's meridian observations
 and the digitized Bordeaux Carte du Ciel plates.  The first observations for the 
M2000 project began in December 1996 and the programme was completed in December 2000 with a median number of 
7 observations per star. Unfortunatly, the  digitization of 
the Carte du Ciel was delayed due to the lack of an available scanner. The plate processing started in 
December 2000 
at the APM measuring machine in Cambridge (UK), and should be finished in July 2001. In this first paper, 
the meridian catalog of 
positions is presented. The Bordeaux meridian 
circle, its CCD drift scan camera and image processing 
are briefly presented in Sect. 2. Sect. 3 concerns the global reduction technique in astrometry and
photometry. Sect. 4
presents some tests performed in order to evaluate the internal and external errors of the catalog. 
Sect. 5 describes the content of the catalog.

\section{Observations and data processing}

\subsection{Instrument and image processing}

The Bordeaux CCD meridian circle has been extensively described in Viateau et
al (\cite{via99}). Here we briefly recall the main characteristics of the instrument and image processing. 

The Bordeaux meridian circle is a refractor with a Texereau objective (202 mm diameter front lens 
and 2368 mm focal length). The detector is a front illuminated 1K CCD Thomson 7896M with 
$19\, \mu m$ pixels corresponding to 
a 28 \arcmin\,
field in declination with a scale of 1.65\arcsec\, per pixel. Its dark current is lower than the 
sky background of the Bordeaux city. Two combined filters 
(GC495 and BG38) give a
visual passband. With the spectral response of the detector cell, the resulting band $V_M$ is displaced 
into the red (520 - 680 nm). 
Due to the large bandwith, the effect of chromatic refraction may be significant, as seen by the comparison 
with the
Hipparcos catalog (Sect. 4.2). The instrument works in TDI mode with an
integration time of $112s/\cos(\delta)$. This process enables the reconstruction of numerical images of 
28 \arcmin \,
in declination
and several hours in right ascension. Such images are automatically processed on-line to give a list of detected
objects. First, the sky background is estimated using a median filter, and subtracted from the image.  
Then objects are defined by at least 2 consecutive pixels above a $3\sigma$  threshold. The position and flux 
of each
detected object are estimated by fitting a two dimensional gaussian flux distribution on the associated 
pixels.
 In case of multiple objects, positions and flux are measured independently if the separation is larger
than 5\arcsec. The dynamical range for the CCD detector is larger than 7 mag.  In the declination range of the 
M2000 catalog, the detection limit is about $V_M=16.5$. Objects brighter than $V = 9.5$ are most often
saturated, with a systematic effect in declination as a
consequence. This effect was easily modeled by fitting polynomials on (Hipparcos minus M2000)  residuals, and  
all saturated stars were systematically corrected in declination.

\subsection{Observationnal strategy}

 In order to observe the whole zone of the Bordeaux Carte du Ciel, $0h \leq \alpha < 24h, 
+11\degr \leq \delta \leq +18\degr$, 
39 strips of 28 \arcmin $\times$ 24h were considered,  with their centers
in declination separated 
by 13 \arcmin. Thanks to this overlap, the 39 strips could be observed 3 times in order to get 6
observations per star plus a thin zone of 2 \arcmin\, where stars were observed 9 times. The strips had their  
length in
right ascension varying from 1 h to 12h depending on the other programs and weather conditions. The
observations for the M2000 program have started in December 1996 and ended in December 2000, after 3306
hours of observations in a fully automatic mode. The objective to get at least 6 observations per star
for more than 95\% of the catalog was successfully achieved. The last year of observation was used to 
fill some gaps in the survey where the
faintest stars could not be observed 6 times due to poor weather conditions.

\section{Reduction procedure}

\subsection{Global astrometric reduction}
As was explained above, the merdian circle observed strips with the same declination width (28 \arcmin) and 
variable lengths in right ascension. The strips are parallel in true equatorial coordinates, with overlaps in both 
directions. For a given
observation, the true absissa of a star image is written :

\begin{equation}
\label{eq_x}
 x=x_m+\epsilon_m+S_m^x(t) 
\end{equation}

\noindent where $x_m$ is the measured value of $x$ and $\epsilon_m$ its measurement error. 
The quantity $S_m^x(t)$ stands for 
correlated departures in position measurements of neighbouring stars, mainly 
attributed to slow variations of the residual angle-of-arrival due to long-period atmospheric fluctuations.
Variations faster than one minute or so are averaged in the TDI mode, their residuals being included
in $\epsilon_m$. 
As explained by Viateau et al. (1999), the deterministic function $S$ has been taken as a B-spline 
which attenuates the oscillations caused by these fluctuations.
For a given strip, or part of a strip, it is written in terms of a set of fitted 
parameters: $S_m^x(t,\lambda_1,\lambda_2,...,\lambda_r$).\\

A similar equation can be written in y :

\begin{equation}
\label{eq_y}
 y=y_m+\epsilon_m'+S_m^y(t) 
\end{equation}

 The relation between the true coordinate $\alpha$ of a star and its image abscissa $x$ on a strip is :

\begin{eqnarray}
\label{eq_a}
\alpha = x+a_0+a_1x+a_2y 
\end{eqnarray}

\noindent where $a_{0}$, $a_{1}$ and $a_{2}$ are the linear model constants 
or "instrumental constants" associated 
with a given strip.  
$a_{1}$ stands for a possible departure to the Earth rotation model, or for any linear 
drift in the instrument attitude. $a_{2}$ stands for misalignement of the CCD columns 
relative to the meridian plane. $a_{1}$ and $a_{2}$ are small quantities. \\ 

A similar relation can be written between the true coordinate $\delta$ and 
the image ordinate $y$:

\begin{equation}
\label{eq_d}
\delta = y + b_{0} + b_{1}x + b_{2}y
\end{equation}

\noindent where $b_{1}$ has the same meaning as  $a_{1}$ in the previous relation,  
and $b_{2}$ stands for departure to the nominal focal scale, mainly due to  
differential refraction in the strip width. The linear model (\ref{eq_a})-(\ref{eq_d}) does
not account for possible optical misalignement or tilt of the CCD plane. Any dependence on $x$
(or time) being dealt with the spline functions, higher order terms would take the form $y^n (n\geq2)$.
We have checked the validity of adding quadratic terms and found it to be unsignificant.\\

The relation between a star image coordinates ($x_{m},y_{m}$) in a strip 
and the true right ascension is deduced from equations 
(\ref{eq_x}) and (\ref{eq_a}) :

\begin{eqnarray}
\label{eq_ax}
\alpha- S_m^x(t,\lambda_1,...,\lambda_r)-a_0-a_1x_m-a_2y_m=x_m+\epsilon_m 
\end{eqnarray}

\noindent with a similar equation in declination. These two relations link the measured image 
position to the celestial coordinates of a star by means of the model parameters 
($a_{0}, a_{1}, \ldots, \lambda_{1}, \ldots, \lambda_{r}$) for each strip.  In practice, strips 
of several hours in right ascension are cut into bands of one hour with some overlap in order to avoid 
the propagation of errors in $x$. In this method, all observed stars are used to determine 
the model parameters. For 
1 million stars observed at least 6 times, one has at least 6 millions 
relations similar to (\ref{eq_ax}), and as many in declination.\\

 The solution of these equations are obtained with a global method (Eichhorn \cite{eic60}, Benevides-Soares
\& Teixeira \cite{ben92}) in which the set of equations is considered as a single system 
to be solved by a least-square
procedure. The equations in $\alpha$ can be written in form of a matrix :

\begin{equation}
\label{eq_mat}
A\vec{\alpha}+B\vec{\lambda}+C\vec{a}=\vec{l}
\end{equation}

The unknown vectors $\vec{\alpha},\vec{\lambda},\vec{a}$ are respectively the right ascensions of all the
observed stars,  the coefficients of the model for atmospheric
fluctuations and the linear model constants of the different strips. ${\vec l}$ is the 
set of coordinates measured on the strips. \\

This system of equations is singular as can be seen in equation (\ref{eq_ax}) : the unknowns $\alpha$ and $a_0$ 
appear
through their difference which means that the origin of right ascensions cannot be determined. Consequently,
 the system (\ref{eq_mat}) has an infinity of solutions and some constraints must be imposed to get a unique
solution. We shall see that these constraints correspond to the choice  of the reference system. 

The system (\ref{eq_mat}) is solved by the iterative method of Gauss-Seidel (see Rapaport \& Le Campion \cite{rap90}
and Le Campion et al \cite{lec92} 
for more details). Two reasons have motivated this
choice: (1) this method is well adapted to the structure of the matrix of equations where only "small systems"
have to be solved, (2) mathematical results have proved the convergence of the iterations toward a solution of
the equations.

The successive steps of the process can be described as follows. From any reference catalog (Tycho-2 in the 
present case) a linear fit will give 
a first estimate of the "instrumental constants", and then a first estimate 
of coordinates ${\vec \alpha_{0}}, {\vec \delta_{0}}$ for all the stars. \\

Starting with ${\vec \alpha = \vec \alpha_{0}}$, the system is solved in 
${\vec a}$ and ${\vec \lambda}$ by least square fit:

\begin{equation}
C{\vec a} + B{\vec \lambda} = ({\vec l} - A{\vec \alpha_{0}})
\end{equation}

The solution is then introduced to solve the system in 
$\Delta \alpha = \alpha - \alpha_{0}$ :

\begin{equation}
\label{eq_mat2}
A\Delta\vec{\alpha}+B\vec{\lambda}+C\vec{a}=\vec{l}-A\vec{\alpha_0}
\end{equation}

\noindent and then the process is iterated. These steps and their interpretation are described in 
Le Campion et al (\cite{lec92}) and  Viateau et al
(\cite{via99}). The process  typically converges toward a least square solution in 5 iterations.

A single solution is obtained by imposing a constraint:

\begin{equation}
\Sigma_i(\alpha_{i,M2000}-\alpha_{i,Tycho-2})=0 
\end{equation}
\begin{equation}
\Sigma_i(\delta_{i,M2000}-\delta_{i,Tycho-2})=0 
\end{equation}

 This constraint is equivalent to add a constant to each coordinate, which comes to put the M2000 
catalog in the reference system of Tycho-2, {\it i.e.} the ICRS. From its construction,  the 
final positions  of the M2000 catalog are independent of the
starting catalog, except on the zero point. This has enabled a direct comparison with the Tycho-2
catalog as described in Sect. 4.3.

In what was just described, it must be noted that each star was observed several times at different epochs 
over four years.
Strictly, the proper motions should be introduced in the equations. Except for very high proper motion stars,
the 4 years observations do not enable a correct determination of proper motions. Consequently, we have chosen 
to neglect temporarily the proper motion displacement over the observing period. The proper motion
determination will be possible when data from the scanning of the Carte du Ciel plates will be introduced 
in the global reduction.

\subsection{Photometric reduction}

The photometric reduction was performed simultaneously with the astrometric reduction, stellar magnitudes 
being entered
as a third unknown parameter after $\alpha$ and $\delta$.
The flux $\Phi$ of each object is initially converted into a magnitude using the standard formula :

\begin{equation}
\label{eq_mag}
V_M=V_0-2.5 \log \Phi
\end{equation}

\noindent where $V_0$ is a constant determined by least squares using  Tycho-2 $V_T$ magnitudes. As our $V_M$ 
magnitude system is slightly different from the $V_T$ system, a color term should be introduced in 
Equ. (\ref{eq_mag}). But colors are not available for most stars yet and a more accurate calibration of 
magnitudes is not possible. The color effect reaches 0.3 mag for the bluest stars. 
 In a 
second step 
the variations of the atmospheric transparency are modeled iteratively by fitting a B-spline on the
residuals of each night :

\begin{equation}
V_M=V_0-2.5 \log \Phi + S_\Phi(t,\lambda_1,\lambda_2,...,\lambda_r)
\end{equation}

\section{Quality Control}
\subsection{Internal precision}

It was decided to keep in the M2000 catalog all the objects which were observed at least 3 times. The
observing program was stopped when more than 95\% of the stars detected in the zone had been observed at least 
6 times. Fig. 1 shows
the histogram of the number of observations per star. In several fields corresponding to other
observing programmes, stars have been observed up to 31 times. The median number of observations per star is 7.\\

After the global reduction, the 3 parameters $(\alpha,\delta,V_M)$ could be
recovered for each star at the different epochs of observations, and their standard errors computed.
The median standard error is a good estimate of the internal precision of the catalog.  As shown in Fig. 2,
standard errors are larger for stars brigther than $V=9.5$ due to saturation, and for the faintest 
stars of the catalog.
 The best results are obtained in right ascension where
standard errors are below 25 mas in the magnitude range $9.5 \leq V_M \leq 13.5$. In the same magnitude interval, the
standard errors in declination and magnitude are lower than 30 mas and 0.03 magnitude respectively. 
In Fig. 3, the histograms of standard errors in $\alpha$, $\delta$ and $V_M$ are presented 
for the whole catalog and for stars brighter than 15. 
The modes and medians of these 6 histograms are summarized in Table 1. \\

It is 
worth noticing that the standard errors in right ascension and declination are slightly increased by the
fact that no attempt was made to calculate the proper motions in the astrometric reduction. As a matter 
of fact,
during 4 years of observations the contribution of the neglected proper motion to the standard error can
reach several mas. The contribution of proper motions in the dispersion of the M2000 measurements is not
easy to estimate as the observations are not uniformly distributed along the time baseline from one strip 
to another.  The typical time baseline between the first and last observation of a star 
is 3 years. The spacing between 2 consecutive observations has 2 typical values : several days
 and about 1 year. 
If $\sigma_\mu$ 
is the standard deviation of the proper motion distribution of all M2000 stars, 
then the maximal contribution of the neglected proper motions is $2\sigma_\mu$ in the
unfavourable and unprobable case of observations only at the beginning and end of the 4 years baseline. Unfortunatly
$\sigma_\mu$ is not known yet. From bright nearby Hipparcos stars, $\sigma_\mu$  was estimated to be 24 $mas.yr^{-1}$ 
for $\mu_\alpha$ and 16 $mas.yr^{-1}$ for $\mu_\delta$.  To get an idea of the value at fainter magnitudes, 
the typical 
stellar content of the M2000 catalog has been simulated 
with the Besan\c con model of stellar population synthesis of the Galaxy (Bienaym\'e et al. \cite{bie87}; 
Robin et al. \cite{rob96}) in 2 representative fields at
high and low galactic latitudes. This simulation showed that the standard deviation of the proper motion 
distribution
reaches 18 $mas.yr^{-1}$ at high galactic latitude. We conclude that the contribution of  
the neglected proper motions to the standard errors in position is significant, especially for the bright part of 
the catalogue ($V < 15$) where random errors  on individual positions have a similar magnitude than 
the dispersion due to proper motions. We expect the median standard errors 
in equatorial coordinates to improve significantly for bright stars  ($\sigma_{\alpha} \simeq 30\, mas$ for $V < 15$) 
when the proper motions will be determined from old Carte du Ciel plates.


\begin{figure}[t]
\resizebox{\hsize}{!}{\includegraphics{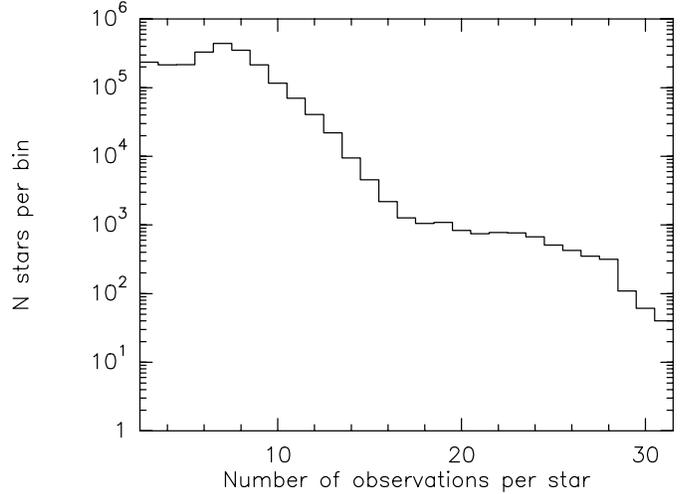}}
\caption{Histogram of the number of observations per star.}
\end{figure}


\begin{figure}[t]
\resizebox{\hsize}{!}{\includegraphics{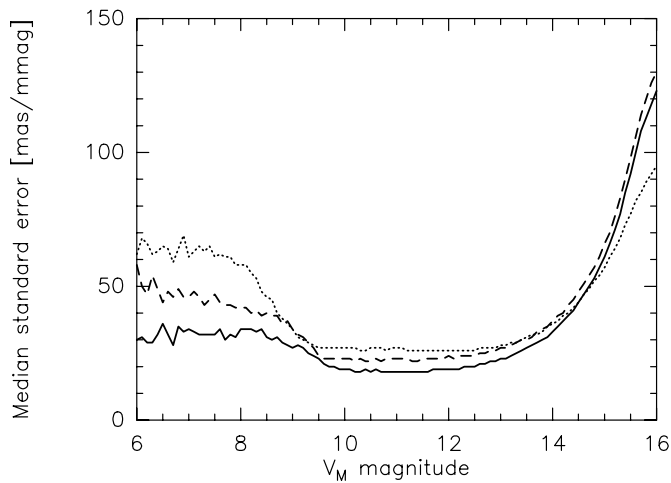}}
\caption{Standard errors in $\alpha$ (full line), in $\delta$ (dashed line) and $V_M$ 
(dotted line) versus magnitude. The median standard errors have
been computed in bins of 0.1 magnitude. }
\end{figure}


\begin{figure}[t]
\resizebox{\hsize}{!}{\includegraphics{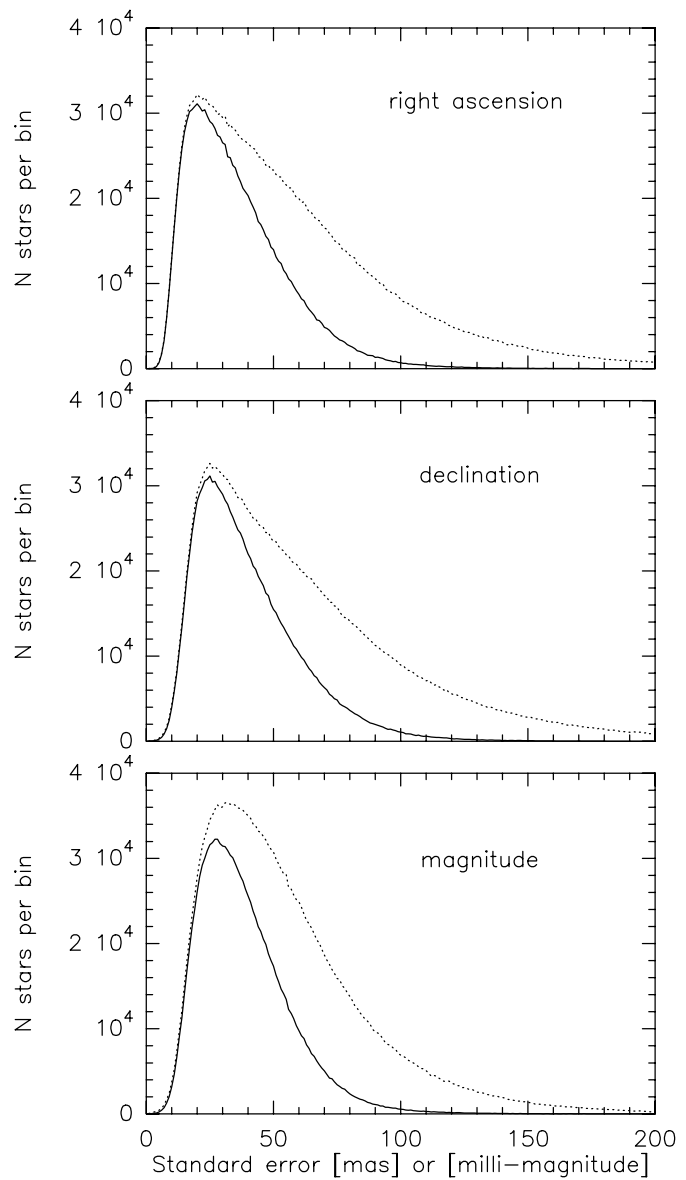}}
\caption{Histogram of standard errors in right ascension, declination and magnitude 
for the whole catalog (dotted lines), and for stars with $V_{M} \leq $15 (full lines).
The bin size is 1 mas, or 1 milli-magnitude.}
\end{figure}


\begin{table}[ht]
\caption[]{Modes and medians of the distribution of standard errors in $\alpha$, $\delta$ and $V_M$ for
the M2000 catalog, with a cut in magnitude and for the whole catalog. The units are mas for $\alpha$ and $\delta$
and milli-magnitudes for $V_M$.}
\begin{flushleft}
\begin{tabular}{| c | c | c | c | c |} \hline
 &\multicolumn{2}{|c|}{$V_M<15$} & \multicolumn{2}{|c|}{full catalog} \\
 &\multicolumn{2}{|c|}{$N\sim 1.2\, 10^6$} & \multicolumn{2}{|c|}{$N\sim 2.3\, 10^6$} \\
\hline
 & mode & median & mode & median \\
\hline
 $\alpha$ & 20 & 31 & 20 & 48 \\
 $\delta$ & 25 & 34 & 25 & 52 \\
 $ V_M  $ & 28 & 34 & 31 & 47 \\
\hline
\end{tabular}
\end{flushleft}
\end{table}

\subsection{External accuracy}

\subsubsection{Comparison with Hipparcos}
The Hipparcos catalog has a positional accuracy better than M2000 by one order of magnitude. 
As a consequence, 
the dispersion of the differences between the two catalogs mainly reflects the external errors of M2000. 
The mean epoch of
the 2 catalogs are different: $1991.25$ for Hipparcos, $\sim 1998$ for M2000.
 To enable a direct comparison, Hipparcos coordinates of each star have been translated to
its mean epoch of observation in M2000. Then the differences in $\alpha, \delta$ have been
computed for the 6613 stars  in common. The mean and dispersion around 
the mean have been computed with an 
iterative $3\sigma$ rejection for
the whole set of stars and for a subset of stars with $V_M>9$, where the M2000
measurements are less affected by saturation. In the $3\sigma$ rejection some 1\% to 3\% of the stars 
were rejected and 
mainly correspond to double
images which were measured as a single object or double object depending on the atmospheric conditions. 
Some double stars
also have poor measurements in the Hipparcos catalog. Most of the time, they are easily 
recognized with their high
standard error in one or both catalogs.

The first look at the results of the comparison  showed that declination measurements were probably 
affected with a 
significant systematic error. The dispersion of the $\delta$ differences was 59 mas for the whole sample
and 46 mas for $V_M>9$,  significantly larger than the dispersion in $\alpha$. The most 
obvious cause is chromatic refraction. As a matter of fact,   a correlation was clearly observed 
between declination departure and colour index of the stars  as seen in Fig. 4. 
A linear fit to the data, eliminating outliers, is : 

\begin{equation}
\label{corbv}
\Delta_\delta = -55.(B-V)_J+44 \,\, \rm{[mas]}
\end{equation}


\begin{figure}[t]
\resizebox{\hsize}{!}{\includegraphics{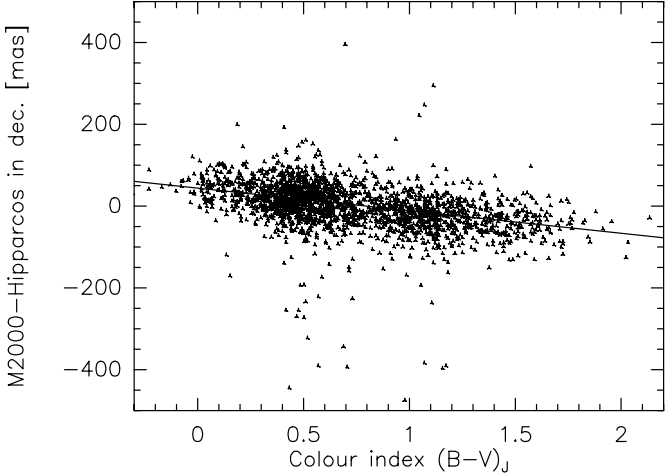}}
\caption{Declination departure for Hipparcos stars versus B-V colour.}
\end{figure}

 The uncertainty on each coefficient is about 1 mas. Due to the refraction dependence on
zenithal distance, we have searched for slope variations within the declination band, without 
significant results. The measured chromatic dependence is in fact larger than the modeled one,
obtained with the instrument transmission, CCD quantum efficiency and blackbody emission spectra
for the observed stars. In the range $0.0 \leq B-V \leq 1.2$, the model slope is found to be -36 mas
at mean zenithal distance of M2000, with variations from -41 to -31 in the declination range 
$+11\degr$ to $+18\degr$. The significant difference with the observed slope in (\ref{corbv})
may indicate additional effects from the instrument optics.
 Unfortunatly the
colour index B-V is now only available for a tiny part of the M2000 catalog ($\sim$ 127\,000 Tycho-2 stars)
where the correction (\ref{corbv}) should be applied. For future developments of M2000 which include 
the proper motion determination, we plan also to cross-identify the M2000 catalog with the 2MASS catalog 
(Skrutskie et al 
\cite{skr97}) in order to estimate the colour index $V_M$-K for each star and apply the adequate correction 
in declination with
this index. The results of the comparison of M2000 with Hipparcos are presented in Table 2. The effect of 
saturation is seen with the higher dispersion in $\delta$ when all Hipparcos stars are considered.


\begin{table}[ht]
\caption[]{Results of the comparison of M2000 with Hipparcos, after the correction (\ref{corbv}) in $\delta$. 
N is the number of stars in common after
the iterative $3\sigma$ rejection, $\Delta$ is the mean difference Hipparcos - M2000 in coordinates (mas), 
$\sigma$ is the standard deviation around the mean.}
\begin{flushleft}
\begin{tabular}{| l | c | c | c |} \hline
 & $N$ & $\Delta$ & $\sigma$ \\
\hline
$\alpha$ & 6372 & -5 & 38 \\
$\delta$ & 6378 & 2 & 46 \\
\hline
$\alpha, V_M>9$ & 2106 & -1 & 37 \\
$\delta, V_M>9$ & 2108 & 3 & 38 \\
\hline
\end{tabular}
\end{flushleft}
\end{table}

In order to estimate exactly the external errors of M2000, the contribution of the errors of Hipparcos 
and their propagation from 1991.25 
to 1998 were computed. We found the errors of the 6613 Hipparcos stars in the M2000 field 
to be on
average 9 mas and 7 mas (medians) respectively in $\alpha$ and $\delta$ at epoch 1998 
(12 mas and 10 mas for $V_M>9$). As the M2000
measurements of bright stars are clearly affected by saturation, from Hipparcos stars with $V_M>9$, 
the external errors of M2000 are estimated to be  35 mas and 37 mas in $\alpha$ and $\delta$ respectively.

\subsubsection{Comparison with Tycho-2}
Tycho-2 was used as the starting point of the iterative global reduction method. 
As explained in Sect. 3.1 the final M2000 catalog is independent, except on the zero point, of the 
reference catalog which was used to initialize the reduction. It is then possible 
to compare directly M2000 with Tycho-2. M2000 has also been compared with a subsample of Tycho-2 made of 
Tycho-1 stars (ESA \cite{esa97}) because of its higher quality.  In order to check the quality of the faint part
of Tycho-2, we have also compared M2000 with a subsample made of Tycho-2 stars not included in Tycho-1. 
The limiting magnitude of Tycho-1 is about 11.5,
whereas it reaches 12.5 for Tycho-2.
  As for Hipparcos,
an iterative 3$\sigma$ rejection was necessary to eliminate from the comparison spurious measurements mainly due 
to multiple stars.  As B-V colours were available for all these stars, we tried to estimate the chromatic 
refraction from Tycho-2 measurements. Due to large random errors, this could not be done in a satisfactory way, but 
restrincting the dataset to Tycho-1 stars, the correction was found to be similar, within the error bars, to
Equ. (\ref{corbv}) obtained with Hipparcos and it was applied to the declinations. The results are presented 
in Table 3, where are also given the Tycho-2 median standard errors ($\sigma_T$, model-based) for 
the comparison stars. The standard deviations ($\sigma$) presented in Table 3 result from the convolution of 
the external errors of Tycho-2 ($\sigma_T$) and M2000 ($\sigma_{M2000}$), so we can write :

\begin{equation} 
\label{accuracy}
\sigma^2=\sigma_{M2000}^2 + \sigma_T^2
\end{equation}

 In Table 3,  $\sigma_T'$ is an estimate of
$\sigma_T$ deduced from Equ. (\ref{accuracy}), assuming M2000 external errors of 35 mas in $\alpha$ and 37 mas in
$\delta$.


\begin{table}[ht]
\caption[]{Results of the comparison of M2000 with Tycho-2, with a subsample of Tycho-2 made of Tycho-1 stars
and with the complementary sample made of the Tycho-2 stars which were not included in Tycho-1 and which 
represents the faint part of Tycho-2. 
The columns $\Delta$ and $\sigma$ are the same as in Table 2. $\sigma_T$ is the Tycho-2 median standard error of
the comparison stars.  $\sigma_T'$ is an estimate of
$\sigma_T$ assuming M2000 external errors of 35 mas in $\alpha$ and 37 mas in
$\delta$. 
Chromatic effects in $\delta$ have been corrected.}
\begin{flushleft}
\begin{tabular}{| l | c | c | c | c | c |} \hline
 & $N$ & $\Delta$ & $\sigma$ & $\sigma_T$ & $\sigma_T'$ \\
\hline
$\alpha$, Tycho-1 & 49476 & -1 & 48 & 30 & 33\\
$\delta$, Tycho-1 & 49657 & 4 & 48 & 34 & 31\\
\hline
$\alpha$, Tycho-2 & 120645 & 0 & 87 & 56 & 80 \\
$\delta$, Tycho-2 & 122210 & 2 & 79 & 67 & 70 \\
\hline
$\alpha$, Tycho-2 faint & 73673 & 1 & 121 & 82 & 116 \\
$\delta$, Tycho-2 faint & 74486 & 0 & 105 & 101 & 98 \\
\hline
\end{tabular}
\end{flushleft}
\end{table}

 	There is no systematic difference between the right ascensions of M2000 and Tycho-2, while there is 
an offset of 2 mas in $\delta$ resulting from the correction of refraction. In Table 3, $\sigma_T$ has to be 
compared with $\sigma_T'$. They are in a very good agreement in both coordinates for the sample made of Tycho-1 stars.
The agreement is also satisfactory in $\delta$ for the 2 other samples. But a large difference is observed in
$\alpha$ between the model-based external errors of Tycho-2 and their estimates with M2000. 
 External errors of M2000 were estimated 
in the magnitude range [9 -- 10.5] and in sect. 4.1 we have shown that M2000
measurements have internal errors lower than 30 mas in the magnitude range [9.5 -- 13.5]. We thus exclude M2000
to be responsible for the resulting high dispersion when compared with Tycho-2.
We conclude that the model-based external errors of Tycho-2 were underestimated in $\alpha$ for the faintest part
of the catalog.  

The comparison of M2000 with Tycho-1 stars from Tycho-2, has also provided a rough estimate of the external 
error in photometry, 0.13 mag, which
is not very significant because the bandpass of the 2 instruments are not exactly the same. The dispersion of 
the residuals in magnitude reaches 0.20 mag when all Tycho-2 stars are considered, and 0.26 if the faint part 
of Tycho-2 is considered.

\section{Content of the catalog}

\subsection{Stellar content}
The catalog is made of 2 275 933 objects measured at least 3 times. Fig. 5 shows the histogram of these 
objects in $V_M$. 
The limiting magnitude is $V_M=16.3$. From the shape of the histogram, the limit of completness is estimated 
to be $V_M=15.4$. The field of M2000 covers 2440 square degrees, i.e. 6\% of 
the whole celestial sphere. With its sky coverage, limiting magnitude and astrometric accuracy,
M2000 is already a valuable sample to test models of the stellar content of our Galaxy from star counts.
 In the near future, proper motions and colours 
will become available. A project has already started to take the census of high proper motion
stars in this area.
Fig. 6 is a map of density which shows the whole Bordeaux zone, with stars brighter than $V = 15.4$. The disc of 
the Galaxy is observed at mean galactic longitudes 
$\ell \simeq 45\degr$ and $\ell \simeq 200\degr$. Irregular bands of dust are clearly identified by the deep
interstellar absorption through the disc. The open cluster, M67 (08h 50.4mn +11\degr\, 49\arcmin) is clearly seen
as a peak of density, as well as two close open clusters, NGC 1817 (05h 12.1mn      +16\degr \, 42\arcmin)
and NGC 1807 (05h 10.7mn      +16\degr 32 \arcmin).


\begin{figure}[t]
\resizebox{\hsize}{!}{\includegraphics{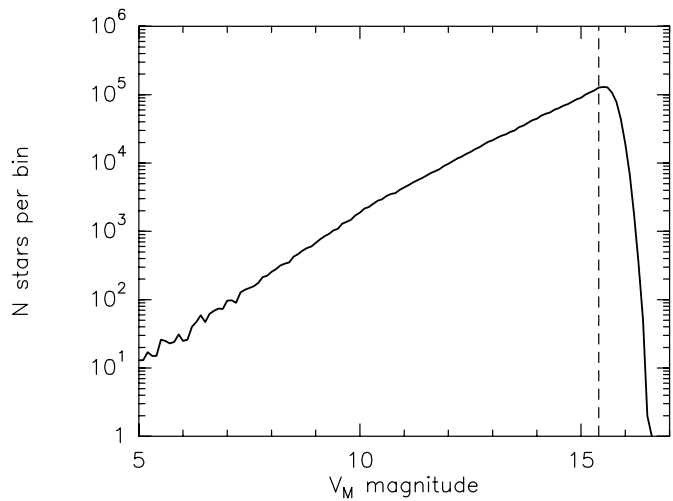}}
\caption{Histogram of the $V_M$ magnitudes. The bins are of 0.1 magnitude. The vertical line at $V_M=15.4$ shows
the limit of completness of the M2000 catalog.}
\end{figure}


\begin{figure*}[t]
\resizebox{\hsize}{!}{\includegraphics{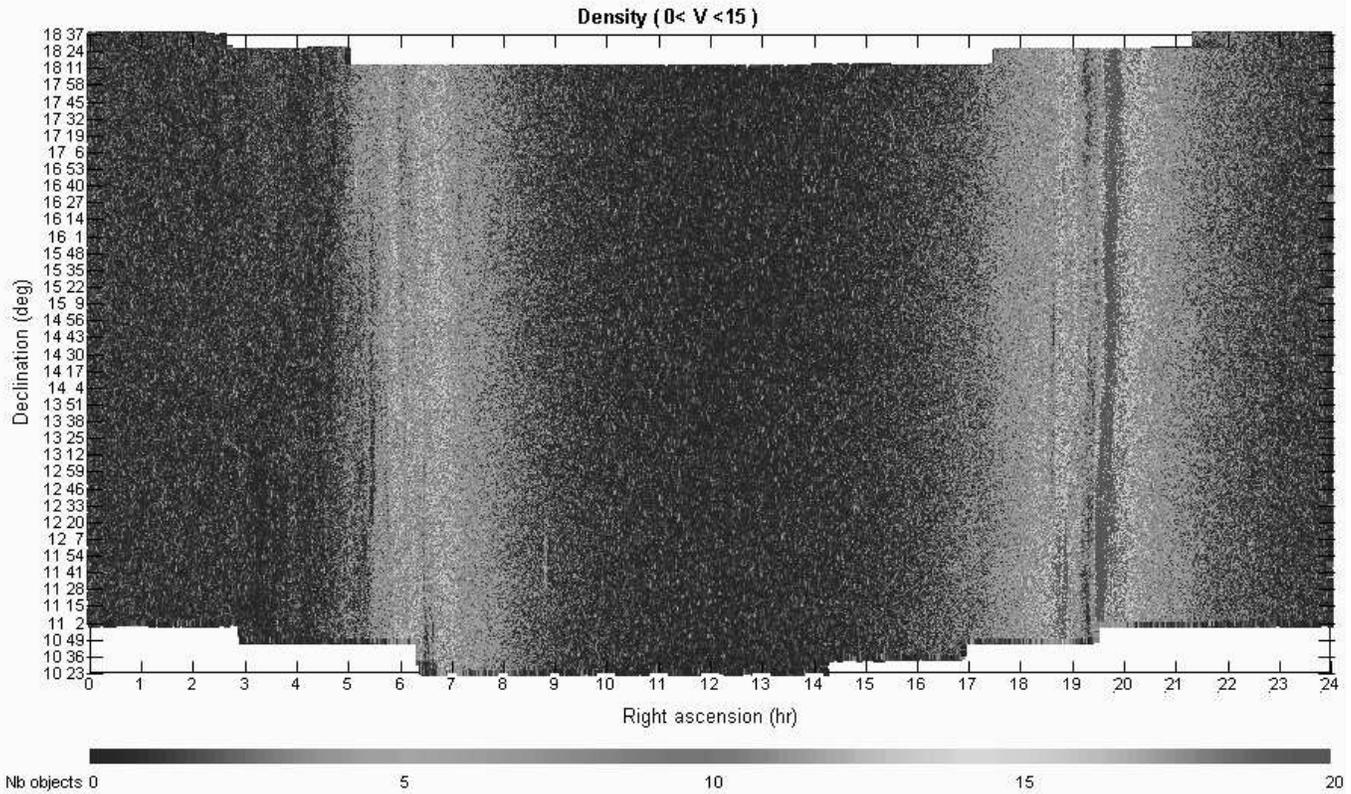}}
\caption{The map of density of objects with $V \leq 15.4$. The grey scale indicates the number of objects 
in pixels of 105s $\times$ 70\arcsec.}
\end{figure*}

\subsection{Format of the catalog}
The catalog is distributed in the form of 24  ASCII files corresponding to one hour intervals in 
right ascension. A system of identification has been elaborated which relies on the strips of observations. 
Tycho-2 stars are identified by their catalog number. An attempt to identify high proper motion
stars from the catalogs NLTT (Luyten \cite{luy79}) and Lowell (Giclas et al. \cite{gic71}), and
galaxies from the catalog RC3 (de Vaucouleurs et al. \cite{vau91}) has been made but it is incomplete due to 
frequent unaccurate coordinates in these catalogs. For each M2000 object, are given the J2000 
coordinates at the mean epoch of observations, the $V_M$ magnitude, the standard errors of $(\alpha, \delta,
V_M)$, the number of used observations in each variable, the mean epochs of observation in $\alpha$ 
and $\delta$ (they can differ due to the $3\sigma$ elimination which treats separately the coordinates).
The M2000 catalog is available at the CDS but due to its large size a CD-ROM can be ordered through
the website http://www.observ.u-bordeaux.fr/$\sim$soubiran/m2000.htm.

\section{Conclusion}
We have presented M2000, an astrometric  catalog in the zone $+11\degr \leq \delta \leq
+18\degr$. M2000 provides J2000 coordinates and $V_M$ magnitudes for $\sim 2.3\, 10^6$ objects which have 
been observed at least 3 times, 7 times on average and up to 31 times. Coordinates $\alpha$ 
and $\delta$ have an internal precision  (median standard error) of 31 and 34 mas for $V < 15$ 
($\sim 1.2\, 10^6$ 
objects), and 48 and 52 mas for the whole catalog. The precision of the $V_M$ magnitudes is $\sim$ 40 
millimag.

From comparison with Hipparcos and the bright part of Tycho-2, we have estimated 
the external errors to be 35 in $\alpha$ and 37 mas in $\delta$. We are aware of a significant
chromatic effect in $\delta$, partly due to the atmospheric refraction. External errors reach 
50 mas in $\delta$ if a correction is not applied. This correction will be possible very soon 
for all stars when the 
cross-identification with the near-infrared survey 2MASS will be achieved and colours available for
the whole catalog.

We have shown that the model based 
standard errors of the faint part of the Tycho-2 catalog have probably been underestimated.

The measurement of proper motions is now under progress with old Carte du Ciel plates. We expect to obtain 
a precison of 2 $mas.yr^{-1}$. With colours and accurate proper motions on 2440 square degrees, with
a limit of completness of $V_M=15.4$, M2000  
will be a major dataset for galactic structure studies. 

\begin{acknowledgements}
We are very thankful to R. Teixeira and P. Benevides-Soares who are closely associated to the evolution
of the Bordeaux meridian circle. We aknowledge with gratitude financial support from the CNRS "Programme National de
Physique Stellaire" to maintain the instrument. 

\end{acknowledgements}

\end{document}